\newcommand{\g}{$\gamma$}
\documentclass[fleqn,10pt]{wlscirep}
\usepackage[utf8]{inputenc}
\usepackage[T1]{fontenc}

\title{Towards machine learning aided real-time range imaging in proton therapy}
\author[1,*]{Jorge Lerendegui-Marco}
\author[1,]{Javier Balibrea-Correa}
\author[1]{V\'ictor Babiano-Su\'arez}
\author[1]{Ion Ladarescu}
\author[1]{C\'esar Domingo-Pardo}
\affil[1]{Instituto de F{\'\i}sica Corpuscular, CSIC-University of Valencia, Valencia, Spain}

\affil[*]{jorge.lerendegui@ific.uv.es}

\begin{abstract}
Compton imaging represents a promising technique for range verification in proton therapy treatments. In this work, we report on the advantageous aspects of the i-TED detector for proton-range monitoring, based on the results of the first Monte Carlo study of its applicability to this field. i-TED is an array of Compton cameras, that have been specifically designed for neutron-capture nuclear physics experiments, which are characterized by \g-ray energies spanning up to 5-6 MeV, rather low \g-ray emission yields and very intense neutron induced \g-ray backgrounds. Our developments to cope with these three aspects are concomitant with those required in the field of hadron therapy, especially in terms of high efficiency for real-time monitoring, low sensitivity to neutron backgrounds and reliable performance at the high \g-ray energies. We find that signal-to-background ratios can be appreciably improved with i-TED thanks to its light-weight design and the low neutron-capture cross sections of its LaCl$_{3}$ crystals, when compared to other similar systems based on LYSO, CdZnTe or LaBr$_{3}$. Its high time-resolution (CRT$\sim$500~ps) represents an additional advantage for background suppression when operated in pulsed HT mode. Each i-TED Compton module features two detection planes of very large LaCl$_{3}$ monolithic crystals, thereby achieving a high efficiency in coincidence of 0.2\% for a point-like 1~MeV \g-ray source at 5~cm distance. This leads to sufficient statistics for reliable image reconstruction with an array of four i-TED detectors assuming clinical intensities of 10$^{8}$ protons per treatment point. The use of a two-plane design instead of three-planes has been preferred owing to the higher attainable efficiency for double time-coincidences than for threefold events. The loss of full-energy events for high energy \g-rays is compensated by means of Machine-Learning based algorithms, which allow one to enhance the signal-to-total ratio up to a factor of 2.
\end{abstract}
\begin{document}

\flushbottom
\maketitle
%
%
\thispagestyle{empty}

\section*{Introduction}
Proton therapy in comparison to conventional radiation therapy is able to target the tumor thanks to the maximum dose deposition at the end of the trajectory of the protons (Bragg peak) and its finite penetration in matter. As the dose deposit beyond this distal edge is very low, proton therapy minimizes the damage to neighbouring tissues compared to photon therapy and is hence especially well-suited for tumors close to sensitive organs and in pediatric cases because the lower dose received by healthy tissues reduces the long-term secondary effects~\cite{Knopf:13}. However, inherent range uncertainties associated to anatomical changes, patient setup errors and range errors from uncertainties in particle stopping power, and imaging reconstruction artifacts~\cite{Kraan:2015} lead to the application of conservative safety margins. Indeed, up to 1~cm of margin is considered nowadays for a prescribed range of 30 cm~\cite{Paganetti:12}, limiting significantly the potential benefits of protons over photons. 

In this context, several experimental methods to verify the proton beam range have been developed in recent years, mainly based on the monitoring of the secondary \g-rays, neutrons or positron emitters produced in nuclear reactions along the proton trajectory. Prompt gamma (PG) monitoring  allows a proper assessment of the distal dose falloff of the proton beam during the treatment~\cite{Krimmer:18}. Unlike the produced positron emitters~\cite{Moteabbed:11} or neutrons~\cite{Ytre-Hauge:19}, the spatial distribution of the emitted prompt gamma rays shows a very close correlation with the proton dose distribution at the end of the beam~\cite{Min:06}. Moreover, these gamma rays are prompt, which means that they are emitted within 1~ns after the collision, which is key for the online verification of the proton range. 

In the last decade, several research groups have designed and tested PG monitoring systems based in prompt gamma timing (PGT)~\cite{Golnik:14,Hueso:15}, \g-ray spectroscopy~\cite{Verburg:14,Hueso:18}, or prompt gamma imaging (PGI) methods. The latter are based on either passive collimation, such as knife-edge-slit cameras~\cite{Smeets:12,Perali:14,Priegnitz:15}, or in active collimation, where most efforts are focused in the development of Compton cameras~\cite{Everett:1977}. Compton cameras are electronically collimated imagers, which represent a promising solution for the imaging of gamma-rays of a few MeV. These systems, in contrast to passive collimated cameras, present higher efficiency and expanded field-of-views which allow reconstructing two- (2D) or even three-dimensional (3D) images instead of one-dimensional (1D) profiles~\cite{Krimmer:18,Hueso:16}. However, there are several challenges that need to be addressed for a reliable implementation of this methodology in the clinical case~\cite{Wronska:20}. Indeed, in-vivo range monitoring remains still an issue for most of the Compton cameras under development~\cite{Ortega:15,Polf:15,McCleskey:15,Hueso:16,Golnik:16,Taya:16,Aldawood:17,Koide:18,Llosa:16,Draeger:18,Parajuli:19,Fontana:20,RosGarcia:20}. These limitations are related to the limited coincidence efficiency of some of the detectors~\cite{McCleskey:15,Golnik:16,Llosa:16}, the high counting rates in clinical conditions~\cite{Krimmer:15,Hueso:16, Koide:18}, the spatial resolution~\cite{Taya:16}, the signal-to-background ratio that is challenged by contaminant reactions~\cite{Ortega:15, Rohling:17}, and the CPU processing-time required by the corresponding image-reconstruction algorithm~\cite{Draeger:18}. 

There are remarkable similarities between most of these challenging effects of PGI in proton therapy, and those encountered in other nuclear physics fields, such as neutron-capture cross-section measurements employing the time-of-flight (TOF) technique~\cite{Guerrero:13}. These similarities are discussed in the following with some detail. In a TOF neutron capture experiment a pulsed beam of neutrons is shot against a small sample, which typically has a small mass of grams or even milligrams. The reaction of interest, radiative capture, leads to the formation of a compound nucleus, which de-excites emitting a prompt cascade of \g-rays. Thus, common \g-ray energies typically span from a few keV up to 5-6~MeV, which is similar to the range of \g-ray energies from proton-induced inelastic reactions in the carbon, oxygen and nitrogen atoms of the human tissues. On the other hand, in neutron-capture TOF experiments the elastic scattering channel dominates versus the radiative channel of interest. These stray neutrons can be captured in the detector itself or in the surrounding materials, thereby enhancing the \g-ray background level and further obscuring the observation of the channel of interest. For the latter reason, an effort is made to design detection systems of high intrinsic \g-ray efficiency and as transparent as possible to neutrons~\cite{Plag03,Balibrea:21,Domingo:20}. This reduces the intrinsic background level and enhances the signal-to-background ratio. Similarly, PGI is also challenged by the background induced by neutrons originating from nuclear reactions of the primary proton beam~\cite{Krimmer:18,Ytre-Hauge:19,Biegun:12}. In the case of carbon-ion beams, neutron production is even more pronounced and their discrimination against prompt \g-rays is an issue~\cite{Krimmer:18}. Therefore, optimization of the detection system in terms of neutron sensitivity becomes also an aspect of interest for hadron therapy, although not much attention has been put into this aspect in recent optimization studies~\cite{Rohling:17,Yao:19}. 

In order to overcome the experimental difficulties discussed above for TOF nuclear experiments, we have developed a total-energy detector with gamma-ray imaging capability, called i-TED~\cite{Domingo:16,Babiano:21}. i-TED is an array of four individual Compton imaging modules, each of them consisting of two position-sensitive detection layers based on large monolithic LaCl$_{3}$(Ce) crystals. A detailed description of the first such i-TED module can be found in Ref.~\cite{Babiano:20}. i-TED features an excellent time resolution for enabling the TOF technique~\cite{Babiano:21} and it has been especially designed to attain a high detection efficiency, a low sensitivity to scattered-neutron backgrounds and a high image resolving power. Its performance can be further enhanced by using innovative methods based on machine learning (ML) techniques~\cite{Balibrea:21,Babiano:21}, which are very powerful thanks to the multi-dimensional structure of the data acquired with i-TED operated in S\&A coincidence, including the energy deposition, position of interaction and time difference of the \g-ray interaction in the two detection planes. 

While the work presented here is entirely based on Monte-Carlo (MC) simulations, the i-TED detector has been already developed~\cite{Domingo:16, Olleros:18, Babiano:19, Babiano:20, Balibrea:21} and employed for TOF  experiments at CERN n\_TOF~\cite{Babiano:21}. Therefore, intrinsic performance parameters such as energy resolution~\cite{Olleros:18}, 3D intrinsic position resolution~\cite{Babiano:19,Balibrea:21} and efficiency~\cite{Babiano:20} have been experimentally validated and are realistically included in the present study. Based on our previous extensive experience in MC simulations of detectors~\cite{Domingo:16,Babiano:21} and neutron production and transport~\cite{Lerendegui:16_2}, in the present simulation work we have been able to account for both geometrical and physical effects to a great level of detail.

Although most of the work carried out so far with i-TED in terms of neutron-capture experiments has been based on an adaptation of the back-projection (BP) method~\cite{Wilderman:98}, for the present study we have implemented two additional imaging algorithms which are the stochastic origin ensemble (SOE) method~\cite{Andreyev:16}, and the analytical algorithm (AA) of Tomitani et al.~\cite{Tomitani:02}. This has allowed us to better illustrate the benefits of the aforediscussed aspects.

Finally, two other aspects are worth mentioning. On one hand, we have been able to implement a novel event classification algorithm, that allows one to significantly improve the quality of the results by filtering out high-energy \g-ray events with incomplete energy deposition. To the best of our knowledge, this is the first time that such a machine learning (ML) technique is successfully applied to this aim, thereby enhancing the signal-to-background ratio by up to a factor of two. On the other hand, we have been able to boost the time-performance of the image reconstruction algorithms by means of an advanced graphical-processing unit (GPU) implementation, which led to reconstruction times of the order tenths of seconds for the most complex of the implemented algorithms. As discussed below, the latter two aspects in conjunction with the high intrinsic efficiency of the i-TED design, turn out of great interest when aiming at real-time ion-range monitoring.

\section*{Results}\label{sec:Results}
To demonstrate the performance of a detector like i-TED for PGI, a series of MC calculations were carried out. In the simulation, a pencil-beam of 120~MeV protons with a spatial spread of $\sigma$=3~mm and a total intensity of 2$\times$10$^{10}$ protons impinges on water and PMMA phantoms with a size of 10$\times$10$\times$20~cm$^{3}$. The energy of the protons, similar to that of previous works~\cite{Verburg:12, Pinto:16,Krimmer:18,Yao:19}, was chosen to match the proton range in water and PMMA (10.7~cm and 9.3~cm, respectively) with the center of the phantom, where the detectors are pointing to.

\begin{figure}[!htbp]
\centering
\includegraphics[width=0.5\linewidth]{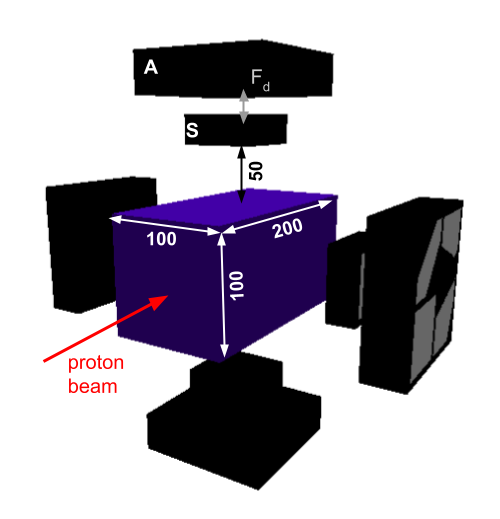}
\caption{Schematic diagram of the simulated geometry, as implemented in the \textsc{Geant4} code. See text for details.}
\label{fig:setup}
\end{figure}

Figure~\ref{fig:setup} shows a schematic view of the simulated geometry. The phantom is surrounded by four i-TED detectors at 50~mm distance from each lateral surface. Each module consists of two layers of LaCl$_{3}$ crystals. The separation between both layers (F$_{d}$ in Fig.~\ref{fig:setup}) can be adjusted for a trade-off between efficiency and resolution~\cite{Babiano:20}. For the present work we use a constant distance of 15~mm. The scatter (S) detector is made from a monolithic block of LaCl$_{3}$ with a size of 50$\times$50$\times$15~mm$^3$. The absorber detector (A) consists of an array of four LaCl$_{3}$ crystals, each one with a size of 50$\times$50$\times$25~mm$^3$. Detector housing, photosensors and other necessary elements have been modeled according to the existing i-TED detector and more details can be found below and in Ref.~\cite{Babiano:20}.

Apart from the geometry itself, an effort was made to implement the most suitable physics libraries for a realistic description of the nuclear reactions both for charged particles and neutrons, and the delivered prompt \g-ray distributions from each isotope. The simulations were carried out with the \textsc{Geant4} toolkit~\cite{Geant4} (v10.6) and the officially released \textsc{QGSP\_INCLXX\_HP} Physics List (PL) was chosen. More details on the impact of the choice of PL are given below.

All the secondary particles generated in the phantom, mainly \g-rays and neutrons, were registered. The largest production yield is found for prompt \g-rays generated in nuclear reactions induced by the 120~MeV proton beam. According to our MC simulations, on average, 0.081 and 0.065 prompt \g-rays (E$>$1~MeV) are produced in Water and PMMA, respectively, by each incident proton. These results are in agreement with the values given in previous works~\cite{Rohling:17, Golnik:14} but might overestimate the actual production~\cite{Pinto:16}. 

\begin{figure}[!htbp]
\centering
\includegraphics[width=0.51\linewidth]{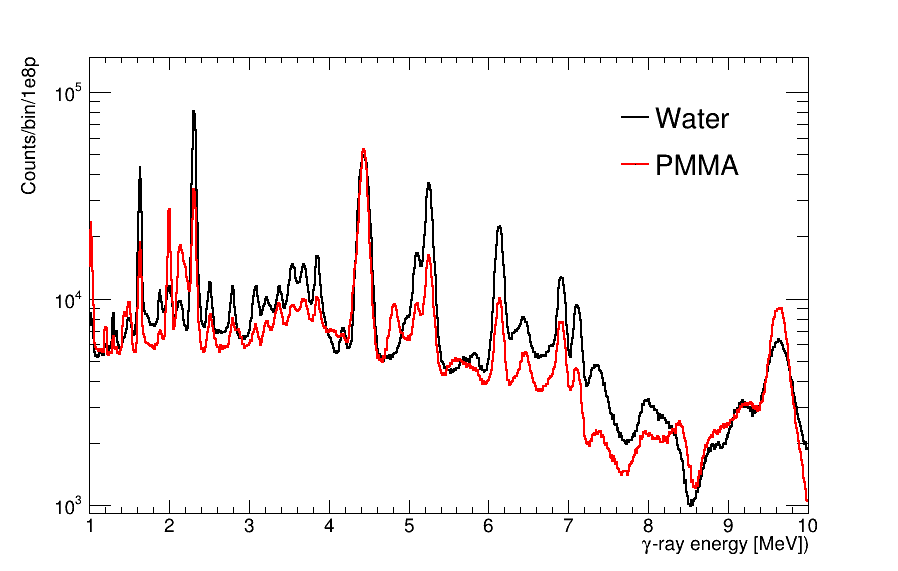}
\includegraphics[width=0.48\linewidth]{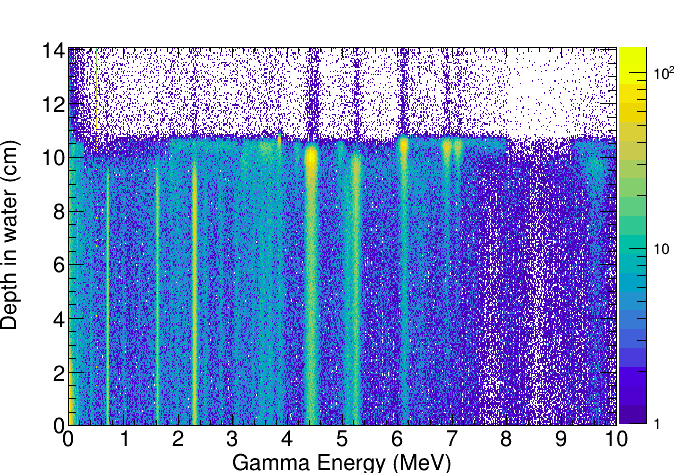}
\caption{Left: Energy spectrum of prompt \g-ray generated by 120 MeV protons along the beam track in water and PMMA. Right: correlation between the \g-ray energy and the emission depth for a water phantom.}
\label{fig:PGSpectra}
\end{figure}

The PG spectra generated by the 120~MeV protons in water and PMMA phantoms, presented in Fig.~\ref{fig:PGSpectra}, are dominated by four \g-ray transitions at 2.3 MeV ($^{14}$N), 4.4 MeV ($^{12}$C), 5.25 MeV ($^{15}$O) and 6.1 MeV ($^{16}$O). More details on the reactions originating these lines can be found elsewhere~\cite{Verburg:14}. The depth distribution of the \g-rays emitted in the irradiation of a water phantom is shown in the right panel Figure~\ref{fig:PGSpectra}. Different emission patterns along the proton track can be identified for each of the lines in the spectrum, associated to the different energy dependence of the underlying nuclear cross sections~\cite{Verburg:12}. Some of the PG lines (for instance 4.4 MeV and 6.1 MeV), show a sharp production maximum in the vicinity of the Bragg Peak, which makes them specially suited for a direct assessment of the proton range. Considering the similarity in terms of PG yield, energy spectra, and proton range in water and PMMA, only the simulations of the water phantom are considered for the results presented hereafter.

To simulate the response of our detection system, all the secondary particles generated inside the phantom are tracked through the geometry model of Fig.~\ref{fig:setup}. For those \g-rays and neutrons interacting with the i-TED detectors, the deposited energy, interaction position and time of all the hits in the S- and A-layers are registered for the subsequent image reconstruction and background assessment. The particle type and energy of the incoming particle were also stored for the analysis described in the following. Experimental effects such as the low energy threshold of 100~keV in each crystal, and the resolution on \g-ray hit position and deposited energy were included in the simulations to account for their impact on the imaging resolution (see Ref.~\cite{Babiano:20}). These experimental resolutions include a 4.5\% \textsc{fwhm} energy spread at 500~keV~\cite{Olleros:18} and 1.5~mm \textsc{fwhm} spatial uncertainty in all three space coordinates for the reconstruction of the \g-ray hit location in each scintillation block~\cite{Babiano:19,Balibrea:21}. 

\subsection*{ML-aided prompt \g-ray imaging}\label{sec:PG_imaging}
The four main PG lines of Fig.~\ref{fig:PGSpectra}, corresponding to de-excitation transitions in $^{14}$N, $^{12}$C and $^{15,16}$O, were selected for image reconstruction. For each \g-ray line, Scatter and Absorber (S\&A) coincidence events were selected and the energy deposited by the \g-ray in the two planes was summed to create the so-called add-back spectrum. A energy window of 150 to 300~keV around the peak energies was selected. In order to reduce the delayed gamma and neutron background associated to the moderation and partial capture of neutrons in the phantom, only events firing the S-layer within 10~ns after the proton pulse were selected. This choice of time-of-flight (TOF) selection and its relation with the neutron sensitivity are discussed in more detail in the following. The applicability of such TOF selection in clinical conditions would depend on the specific time structure of the proton accelerator~\cite{Krimmer:18}.

\begin{figure}[!htbp]
\centering
\includegraphics[width=0.8\linewidth]{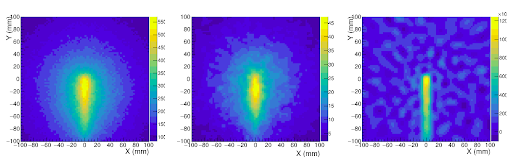}
\includegraphics[width=0.8\linewidth]{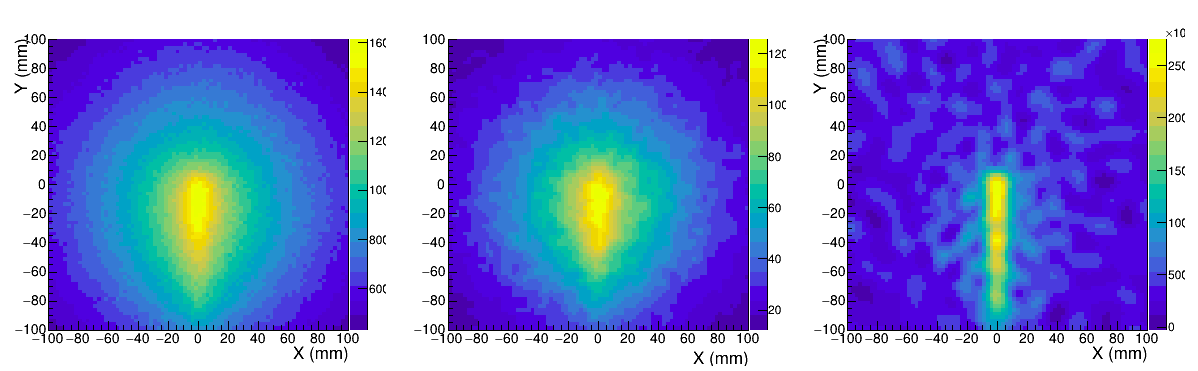}
\includegraphics[width=0.8\linewidth]{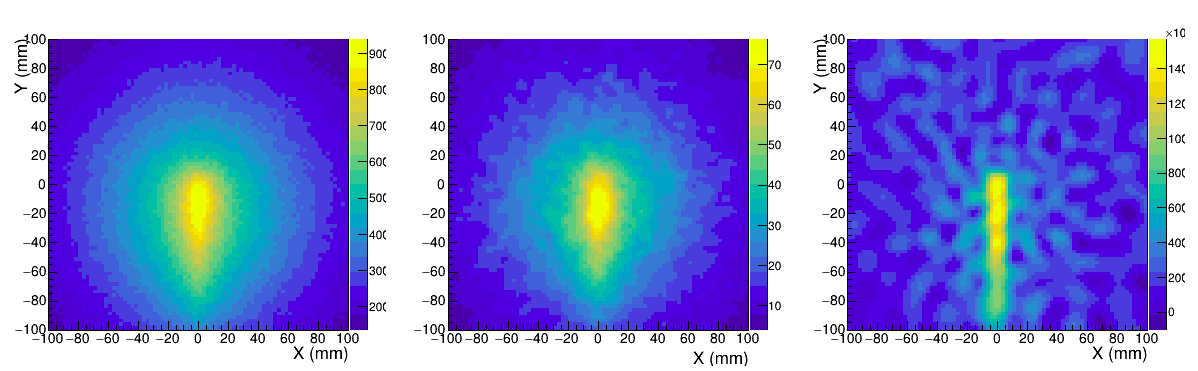}
\caption{From left to right, the imaging algorithms used correspond to BP, SOE and AA, respectively. The top-row figures show the Compton image obtained when only full-energy deposition in S and A detectors are selected. For the middle-row images, no condition on full-energy deposition has been applied and also neutron interactions are considered. The bottom-row images have been obtained considering also all the events but in this case the ML-classifier was used to filter out events with incomplete energy deposition.}
\label{fig:2d all}
\end{figure}

A crucial aspect for the application of PGI to proton range verification is the attainable spatial resolution in the distal edge of the PG depth distribution (see Fig.~\ref{fig:PGSpectra}). Aiming at achieving a good balance between imaging resolution and reconstruction time, three different algorithms have been implemented and tested for the reconstruction of 2D PG images: BP, SOE and AA. More details on the implementation of these algorithms are given in Methods. Fig.~\ref{fig:2d all} shows the Compton images obtained with the three algorithms after selecting in the add-back spectra the four main peaks corresponding to the most intense PG lines. These images have been obtained by combining the statistics of the four i-TED detectors and using the full simulated statistics (2$\times$10$^{10}$ protons) to highlight the attainable spatial resolution and the differences between imaging algorithms.

\begin{figure}[!htbp]
\centering
\includegraphics[width=0.49\linewidth]{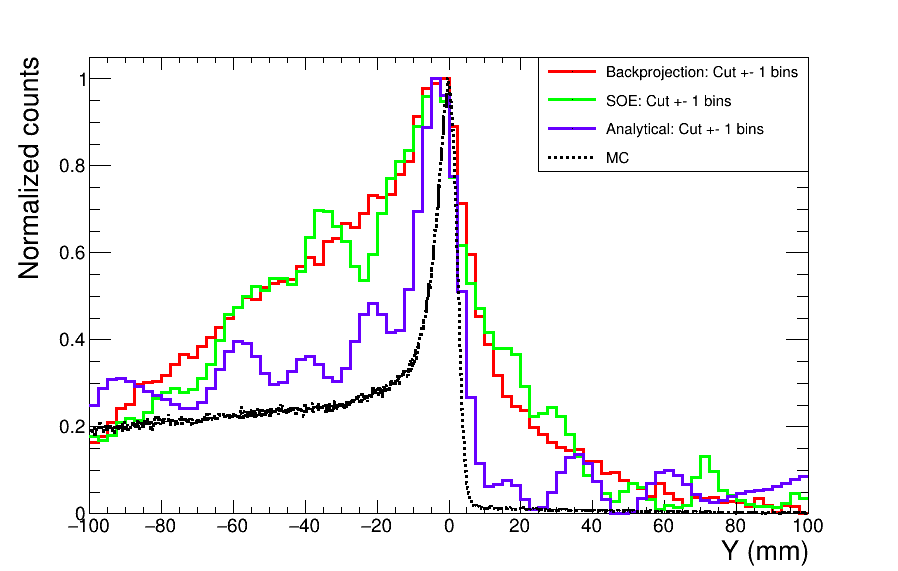}
\includegraphics[width=0.49\linewidth]{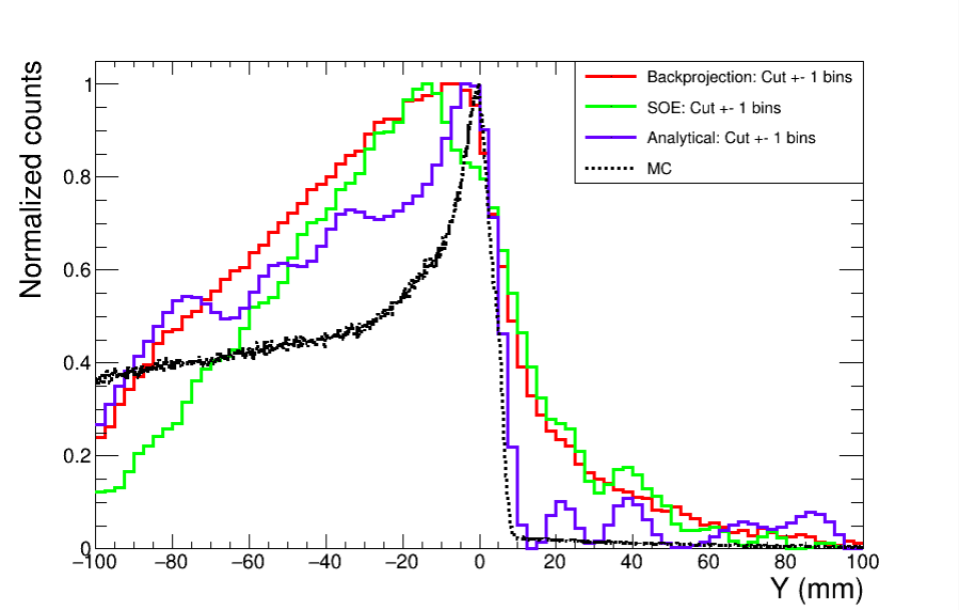}
\caption{1D projection of the Compton image along the proton beam axis obtained using only the 4.4~MeV transition in $^{12}$C (left) and including the four main transitions in $^{14}$N, $^{12}$C and $^{15,16}$O (right). The different solid lines correspond to the BP, SOE and AA reconstruction methods. The true depth distribution (MC) is shown as the black dashed line.}
\label{fig:projectionAlgorithms}
\end{figure}

 If one compares the three images in the upper row of Fig.~\ref{fig:2d all}, obtained using only gamma-ray events with full-energy deposition, the limited resolution and signal-to-background achieved with the simple BP algorithm is only slightly improved with the SOE algorithm (1000 iterations). On the contrary, the analytical approach leads to a clear upgrade in terms of imaging resolution (see Fig.~\ref{fig:2d all}). For a quantitative comparison of the resolution provided by the different algorithms, the projections of the 2D image along the proton beam (Y) axis are presented for different PG energy selections in Fig.~\ref{fig:projectionAlgorithms}. For the case of the $^{12}$C peak, the three algorithms reproduce the position of the PG emission maximum within 5~mm. In both cases ($^{12}$C and 4 main PG lines), the profiles extracted from BP and SOE algorithms start to deviate from the actual fall-off profile below the 80\% of the maximum. On the contrary, the AA method is able to reproduce in a remarkable manner the sharp PG fall-off distribution. Moreover, the AA yields also the best reproduction of the PG depth distribution at shallow depths. The precision in the reconstructed PG emission profiles for the number of protons of clinical interest is discussed later in this work.
 
Besides the finite resolution of the detector and the performance of the imaging algorithm, the hard spectra of the PG lines of interest represents an additional challenge for the satisfactory reconstruction of the PG depth distributions. In 2-plane Compton cameras (CC) like i-TED, the large fraction of \g-ray events with an incomplete energy deposition leads to a deteriorated image reconstruction, as displayed in the central row of Fig.~\ref{fig:2d all}. This limitation has led to the development of 3-plane and multistage CC for applications dealing with high energy gamma-ray transitions~\cite{Llosa:16,Taya:16,McCleskey:15,Draeger:18}. These systems enable a more reliable determination of the \g-ray energy by means of threefold time-coincidences. Nevertheless, such an approach has a significant cost in detection efficiency, which hinders the feasibility of real-time range verification.
\begin{figure}[!htbp]
\centering
\includegraphics[width=0.49\linewidth]{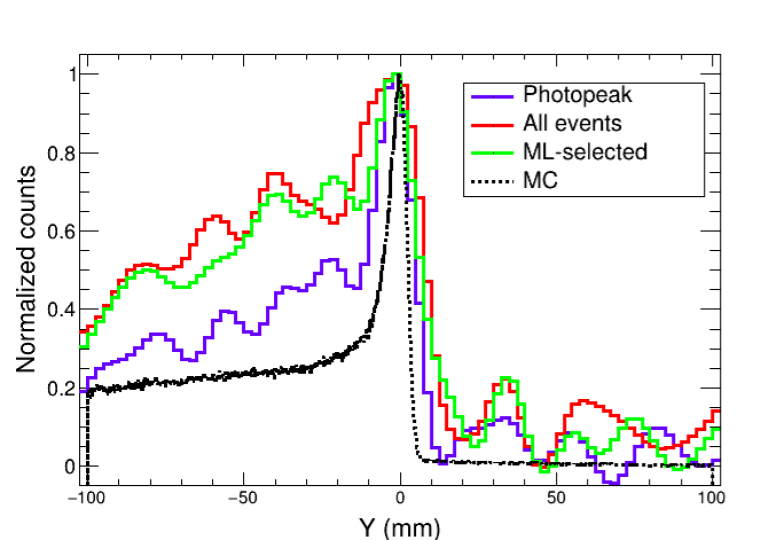}
\includegraphics[width=0.49\linewidth]{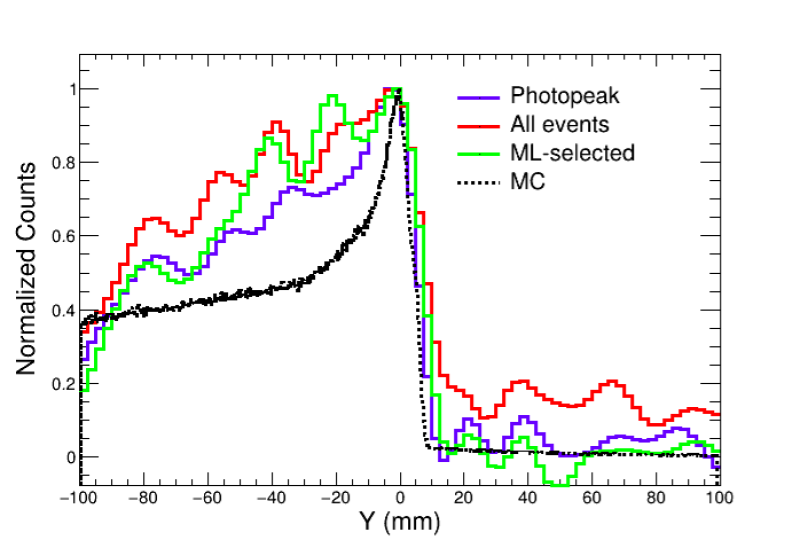}
\caption{1D projection of the Compton images reconstructed with the analytical algorithm using the 4.4~MeV transition in $^{12}$C (left) and combining the four main transitions in $^{14}$N, $^{12}$C and $^{15,16}$O (right). The three solid curves correspond to the ideal image with only full-energy \g-ray events(blue), the image with no selection including neutron events (red) and the corrected result after the ML-classifier is applied (green). The true depth distribution (MC) is shown as the black dashed line.}
\label{fig:image projection}
\end{figure}
 Aiming at keeping the advantage of the high efficiency in i-TED while selecting only full-energy events, an innovative approach based on machine learning (ML) identification of full-energy events has been developed in this work. This ML solution was coded in the \textsc{TensorFlow} deep-learning API~\cite{tensorflow:15} and it has allowed us to enhance the fraction of full-energy events in a factor ranging from 1.5-to-2 in the energy range of interest for PGI (see Fig. ~\ref{fig:MLClassifier} in Methods for the details). As a consequence, to a large extent one can recover the resolution and signal-to-background ratio of an ideal Camera with only full-energy events (see Fig.~\ref{fig:2d all}).

The impact of the ML-aided image reconstruction is shown for the 1D profiles reconstructed with the analytical algorithm in Fig.~\ref{fig:image projection}. This figure shows that the reproduction of the position, width and fall-off of the PG emission profile are improved to some extent thanks to the ML solution (green) with respect to the original reconstruction (red) and getting closer to the ideal reconstruction with only full-energy Compton events (blue). In particular, the deviation between the reconstructed and the true (MC) position of the 4.4 MeV PG fall-off curve (left panel in Fig.~\ref{fig:image projection}), quantified at the 80\% of the maximum, $F_{80\%}$, is reduced from the 3.2 mm in the original reconstruction to only 1.3 mm after the ML selection is applied. As for the depth profile of the 4 main lines combined (right panel), the accuracy improvement in the reconstruction of the PG fall-off ($F_{80\%}$) is more subtle, changing from 3~mm with all the events to 2.5~mm with the ML method.
 
\begin{figure}[!htbp]
\centering
\includegraphics[width=0.45\linewidth]{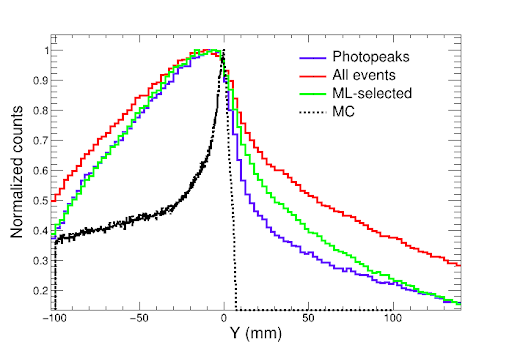}
\includegraphics[width=0.45\linewidth]{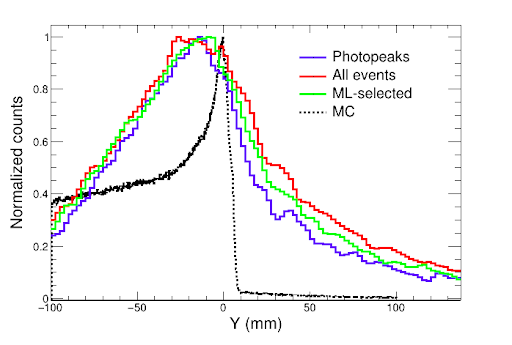}
\caption{1D projection of the Compton images of Fig.~\ref{fig:2d all} along the proton beam axis obtained with the BP (left) and SOE (right) algorithms. The three solid curves correspond to the ideal image with only full-energy \g-ray events (blue), the image with no selection including also neutron events (red) and the corrected result after the ML-classifier is applied (green). The true depth distribution (MC) is shown as the black dashed line.}
\label{fig:MLImpactSOE_BP}
\end{figure}

 The impact of the ML selection in the Compton images is more sizable for the images reconstructed with the faster SOE and BP algorithms, which do not reach the resolution of the aforediscussed AA. The 1D profiles obtained from the BP and SOE images of Fig.~\ref{fig:2d all} are compared in Fig.~\ref{fig:MLImpactSOE_BP} for the same three scenarios. In this figure one can appreciate the noticeable enhancement in peak-to-background ratio, especially for the BP algorithm, related to the application of the ML-based full-energy selection (blue) when compared to the initially reconstructed profile (red).

 \begin{figure}[!htbp]
\centering
\includegraphics[width=0.7\linewidth]{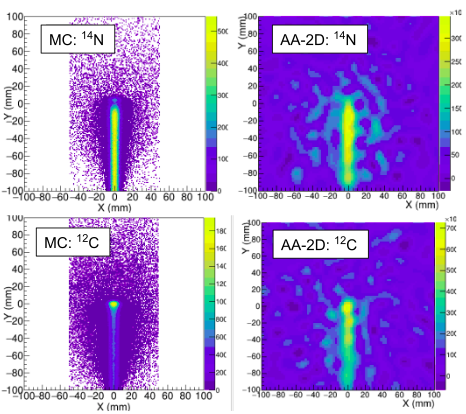}
\caption{2D Compton images obtained with the analytical reconstruction method and the ML-aided event selection (right) compared to the spatial distributions from the MC simulations (left). The top and bottom images correspond to the $^{14}$N peak and $^{12}$C peaks, respectively. }
\label{fig:ImagePeaksMLAA}
\end{figure}
Fig.~\ref{fig:ImagePeaksMLAA} shows some examples of the final results achieved for the imaging of single PG lines using the combination of the AA reconstruction algorithm and the ML-aided selection of full-energy events. The reconstructed 2D images for the 2.3~MeV ($^{14}$N) and the 4.4~MeV ($^{12}$C) lines are compared with the true distribution obtained from the MC simulation, showing a remarkable reproduction of the PG emission distribution and the distal fall-off in both cases.

\subsection*{Impact of neutron-induced backgrounds}\label{sec:n_sensitivity}
Neutrons are produced along with prompt gamma-rays in nuclear reactions during the proton therapy treatment. According to our MC simulations, the prompt neutron yield generated by 120~MeV protons in water is of 0.064 per incident proton, just 20\% smaller than that of \g-rays in the energy region of interest (1-7 MeV). These neutrons are partially moderated before escaping the water phantom and may be captured in the detector sensitive volume or surrounding structural materials, leading to an important source of background. 

The impact of the detector neutron sensitivity was studied by means of simulations of the same set-up and only 10$^{8}$ protons, in which the LaCl$_{3}$ crystals of i-TED were replaced by other inorganic scintillators or semiconductor detectors commonly used in existing or foreseen CC designs. These active detection materials are LYSO~\cite{Rohling:17,Hueso:16}, BGO~\cite{Hueso:16, Krimmer:15,Fontana:20}, CdZnTe~\cite{McCleskey:15,Polf:15,Draeger:18,Yao:19} and LaBr$_{3}$~\cite{Ortega:15,Llosa:16,Aldawood:17,Munoz2021} crystals. Ce:GAGG~\cite{Taya:16,Koide:18} has not been included in the study but its neutron sensitivity could be even higher due to the presence of Gd, featuring one of the largest known thermal cross sections~\cite{Fedorov:20}.

\begin{figure}[!htbp]
\centering
\includegraphics[width=0.51\linewidth]{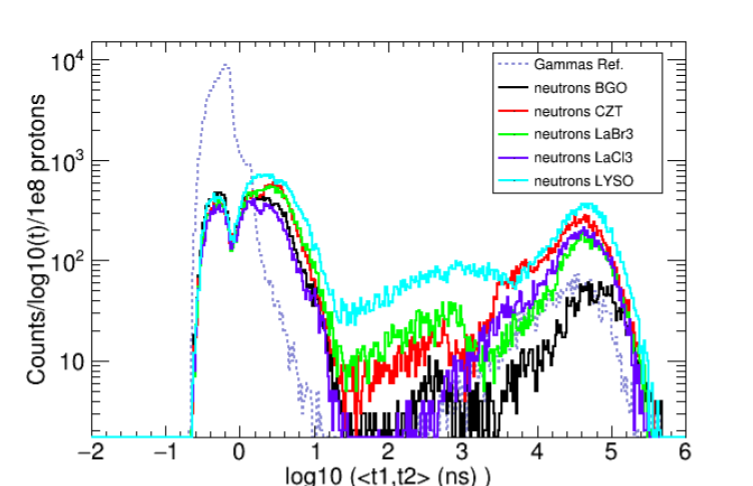}
\includegraphics[width=0.48\linewidth]{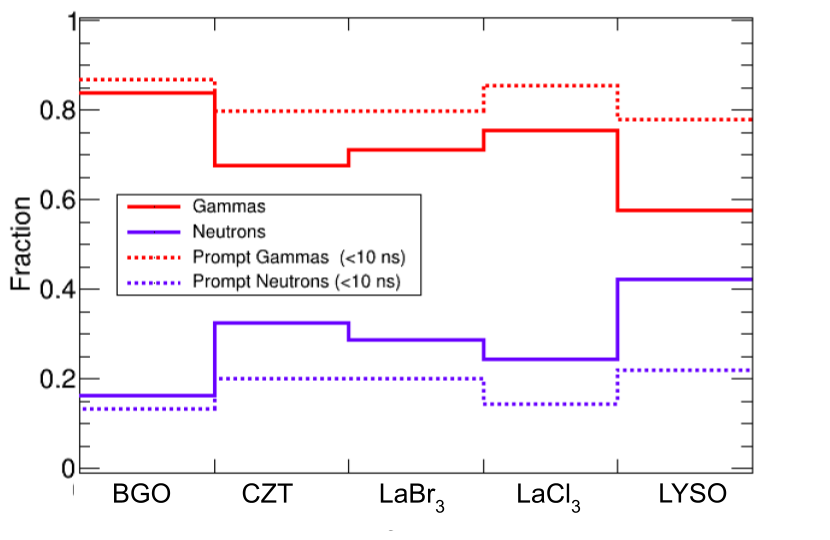}
\caption{Left: Time distribution of the neutron-induced counts registered in a Compton camera (i-TED geometry) for different detection materials. The time distribution of \g-ray counts in LaCl$_{3}$ is shown as a reference in dashed line. Right: Fraction of neutron- and gamma-induced S\&A coincidences for each detector material in the full time window (solid) and after selecting the first 10~ns(dashed).}
\label{fig:nsensitivity}
\end{figure}

Figure~\ref{fig:nsensitivity} shows the time distribution of neutron events in the detector for the different active materials compared to the distribution of \g-ray events. LaCl$_{3}$ shows the smallest efficiency for the detection of prompt neutrons reaching the detector within few ns after the proton bunch, simultaneously with the prompt gamma-rays arising from proton interactions in the water phantom. On the other hand, BGO is the least sensitive material to slow neutrons reaching the detector 1-100$\mu$s after the proton pulse. 

The figure of merit to be studied is the fraction of neutron- and gamma-induced coincidences, displayed in the right panel of Figure~\ref{fig:nsensitivity}. LYSO, a very promising crystal in terms of efficiency and time resolution~\cite{Hueso:16}, shows the highest sensitivity to neutrons, which represent 42\% of the total counting rate. On the other side, apart from its limited energy resolution for Compton imaging, BGO seems the best solution due to its low sensitivity to neutrons. If the clinical accelerator pulse structure allows to set TOF selections of a few ns, as proposed in recent works~\cite{Cambraia:18, Fontana:20}, the contribution of neutron background would be clearly suppressed for all the studied crystals. Our results indicate that after a TOF selection of e.g. 10~ns, LaCl$_{3}$ and BGO would have the lowest fraction (14\%) of neutron coincidences or neutron sensitivity. A complete TOF separation of prompt gammas and neutrons is not feasible if one aims to maximize the \g-ray detection efficiency with a small detector-to-phantom distance. Hence, detectors with a low intrinsic sensitivity  to neutrons, such as the LaCl$_{3}$ based i-TED modules, may represent a valuable solution.

\subsection*{Efficiency and in-vivo real-time PG imaging}\label{sec:efficiency}
 The spatial resolution and low sensitivity to backgrounds attainable with the ML-aided i-TED detector are crucial for its applicability to PGI. Aiming at in-vivo real-time range verification, several additional aspects are of special relevance. The first key feature is the detection efficiency, which has to be high enough to reconstruct PG images with sufficient resolution in clinical conditions. Average clinical beam intensities are of the order of 1-2 nA (i.e. 6-12$\times$10$^{9}$ p/s)~\cite{Krimmer:18} and the relevant clinical scenarios for a single pencil beam in beam-scanning proton RT correspond to the delivery of 10$^{8}$ to 10$^{9}$ protons~\cite{Smeets:12,Hueso:16,Draeger:18}.

Table~\ref{tab:efficiency} summarizes the detection efficiency for the proposed setup composed of four i-TED modules at 100~mm from the proton beam axis displayed in Fig.~\ref{fig:setup}. The resulting efficiency, defined as the number of S\&A coincidences per number of incident protons (as in Ref.~\cite{Yao:19}), is compared for several energy ranges in add-back which comprise single or multiple prompt gamma lines. The impact of the distance between the S- and A-planes, which can be remotely adjusted for an optimum trade-off between efficiency and resolution has been studied. The latter approach refers to the electronic-dynamic collimation implemented in i-TED, which is described in detail in Ref.~\cite{Babiano:20}. The proposed setup of four detectors at 10 cm may not be feasible in some clinical scenarios. In a general case, the efficiency of our imaging system would scale proportional to the number of Compton modules and inversely proportional to the square of the distance.  

\begin{table}[ht]
\centering
\begin{tabular}{|l|c|c|c|}
\hline
                        &   \multicolumn{3}{|c|}{Focal distance (mm)}\\ 
\hline
Energy selection        &  5   &   15  &  30  \\
\hline
All PG (1-7 MeV)        & 2.6$\times$10$^{-4}$& 2.1$\times$10$^{-4}$& 1.6$\times$10$^{-4}$\\
4 main PGs              & 4.3$\times$10$^{-5}$& 3.5$\times$10$^{-5}$ & 2.6$\times$10$^{-5}$\\
$^{12}$C (4.3-4.6 MeV)  & 1.5$\times$10$^{-5}$& 1.2$\times$10$^{-5}$ & 8.8$\times$10$^{-6}$  \\
\hline
\end{tabular}
\caption{\label{tab:efficiency}Detection efficiency for S\&A in time coincidence per incident proton combining the four i-TED modules of Fig~\ref{fig:setup}. Each row corresponds to a different selection in deposited energy and each column shows the result for a different distance between the S- and A-planes. The uncertainties due to counting statistics are below 0.5\%.}
\end{table}

\begin{figure}[!htbp]
\centering
\includegraphics[width=0.45\linewidth]{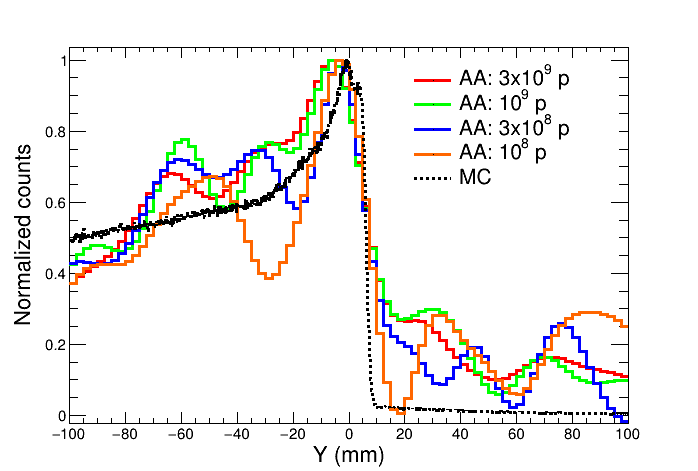}
\includegraphics[width=0.45\linewidth]{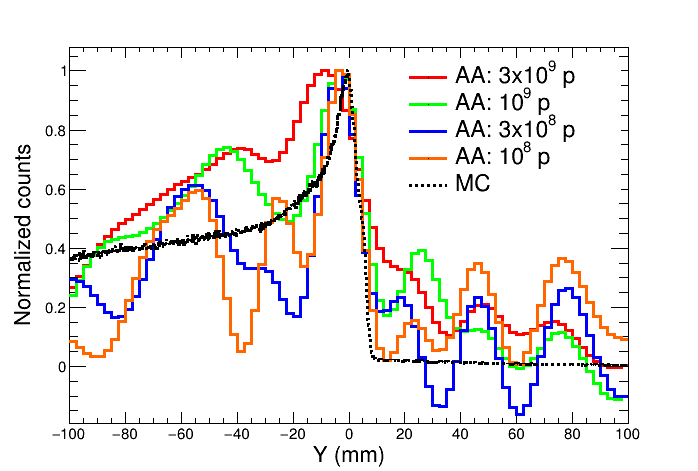}
\caption{1D projection of the Compton images along the proton beam axis obtained the analytical reconstruction algorithm using all the events between 1 and 7~MeV (left) and the 4 main PG lines (right). The solid curves show different proton intensities. The true depth distribution (MC) is shown as the black dashed line.}
\label{fig:ReducedProtonNumber}
\end{figure}

According to the results shown in Table~\ref{tab:efficiency}, only 6\% of the total coincident events are selected with an energy window around the 4.4~MeV peak. If the four main PG lines are combined, as in Fig.~\ref{fig:2d all}, this fraction increases to 16\% of the total coincidences. Hence, in order to achieve sufficient statistics ($\sim$10$^{4}$ events per image) for in-vivo range verification (i.e. $<$10$^{9}$ protons), the combination of several PG lines becomes mandatory. Figure~\ref{fig:ReducedProtonNumber} compares the 1D profiles of the images obtained with the AA algorithm and the ML-based full-energy selection corresponding to different number of protons and two different energy windows. On one hand, selecting the whole energy range from 1 to 7~MeV improves the statistics and thus less fluctuations are obtained in the reconstructed profiles. On the other hand, limiting the energy selection to the four main peaks reduces the background and enhances the fraction of full-energy events, hence improving the resolution in the determination of the proton range. The achieved precision in the reproduction of the PG emission fall-off is quantitatively analyzed in Table~\ref{tab:accuracy}. The values in this table correspond to depth in water taking into account that the reference frame of the Compton images in this work is centered at a depth of 100~mm.  The position of the reconstructed maxima (Max) has been determined from the center of the bin and the positions along the fall-off curve (F$_{90\%}$, F$_{80\%}$ and  F$_{50\%}$) have been linearly interpolated from the histograms in Fig.~\ref{fig:ReducedProtonNumber}. The final reconstructed positions in Table~\ref{tab:accuracy} have been calculated as the average of two independent sets of simulated data and the uncertainties correspond to the standard deviation between the two independent results.

\begin{table}[ht]
\centering
\begin{tabular}{|l|c|c|c|c|}
\hline
1$<E_{s}+E_{a}<$7~MeV  &  Max   &   F$_{90\%}$  &  F$_{80\%}$ &  F$_{50\%}$ \\
\hline
Real (MC)             & 99.70(5)      &   104.33(1) &    105.13(1)   &   106.19(1)        \\
\hline
3$\times$10$^{9}$ p    & 95.0(18)     &  100.6(7)  &  103.27(15)   &   106.9(12)        \\
10$^{9}$ p             & 95.0(18)     &  100.4(15)   &  102.5(11)   &  107.7(6)    \\
3$\times$10$^{8}$ p    & 95.0(18)     &  99(3) &  101.9(10)    &  107.1(7) \\
10$^{8}$ p             & 93.8(4)     &  100(3)  &  102.0(22)    &  107.0(9)   \\
\hline
4 main PG peaks  &   Max.   &   F$_{90\%}$  &  F$_{80\%}$ &  F$_{50\%}$ \\
\hline
Real (MC)              & 99.70(5)   &   100.96(1) &    101.77(1)   &   104.71(1)   \\
\hline
3$\times$10$^{9}$ p    & 96.3(12)   &  98.5(13)           &     101.0(6)     & 106.7(9)          \\
10$^{9}$ p             & 95.0(18)   &  99.1(20)           &    100.9(20)     & 105.2(18)           \\
3$\times$10$^{8}$ p    & 95.0(18)   &  98.5(13)           &     100.1(23)     & 103.7(15)     \\
10$^{8}$ p             & 95.0(18)   &  98(3)           &    99(4)     &  102(4)       \\
\hline

\end{tabular}
\caption{\label{tab:accuracy} PG maximum (Max) and 90\% (F$_{90\%}$), 80\% (F$_{80\%}$) and 50\% (F$_{50\%}$) fall-off positions corresponding to the profiles of Fig.~\ref{fig:projectionAlgorithms} compared to the real PG emission profile in the MC simulations. The values are given in mm. The values in parentheses correspond to the standard deviation of independent data sets.}
\end{table}

According to the results of Table~\ref{tab:accuracy}, deviations between the reconstructed positions and the actual PG distribution are in the range 0.5~mm to 5~mm for the profile selecting the energy window from 1 to 7~MeV. For the case of the 4 main peaks the agreement is in most cases better than 3~mm. As the number of protons decreases, the uncertainties tend to be slightly enhanced. The best precision is obtained for the reconstructed position of the 50\% fall-off (F$_{50\%}$), which agrees with the real (MC) distribution within less than 1~mm in most cases (see Table.~\ref{tab:accuracy}). On the other hand, the position of the PG maxima are systematically underestimated by at least 3~mm and have a lower accuracy related to the 2.5~mm bin size.

The computing-time performance of the relatively complex Compton imaging algorithms is also critical for the in-vivo range verification via PGI. In this work, three different 2D image reconstruction algorithms have been tested. The reconstruction times with the BP and SOE methods are perfectly compatible with real-time imaging (see Table~\ref{tab:time}). For the latter, good quality images containing a minimum of 2$\times$10$^{4}$ coincident events are reconstructed in few seconds using a single-thread CPU calculation. As for the AA method, which yields the best image resolution (see Figs.~\ref{fig:2d all} and \ref{fig:projectionAlgorithms}), we used a GPU-accelerated CUDA~\cite{CUDA} implementation of the code, which allows reconstructing an image in few tenths of seconds, 120 times faster than with a conventional single-thread CPU based approach. A similar acceleration has been reported for 3D position reconstruction algorithms in our previous work~\cite{Balibrea:21}.

\section*{Discussion and Conclusion}
i-TED is a Compton camera array that has been specifically designed for neutron-capture nuclear physics experiments. Several design aspects of i-TED, such as its high time resolution, high efficiency and relatively low neutron sensitivity may become of interest in order to address some of the current challenges in Prompt Gamma Imaging for range verification in proton therapy treatments. The results presented in this work show the prospects of i-TED concerning detection efficiency, sensitivity to the neutron background and spatial resolution attainable with a combination of ML-based full-energy selection method and state-of-the-art reconstruction algorithms. 

This work has presented a MC study on the applicability of i-TED to range verification, where the attainable image resolution is one of the critical issues. The BP and SOE algorithms yielded a good reproduction of the PG maximum emission point in spite of their limited resolution. Much higher resolution images can be obtained with i-TED using the analytical approach of Ref.~\cite{Tomitani:02}, which provided the best reproduction of the distal fall-off for several selections of PG lines with an accuracy ranging from 1 to 3 mm. To achieve these results, a ML classifier, which improved the fraction of correct Compton events up to a factor 2, has been developed to partially compensate the deterioration of the images due to partial-energy events.

In comparison, the recent MC study by Yao et al.~\cite{Yao:19}, obtained accuracies within 2~mm for the same F$_{80\%}$ and F$_{50\%}$ magnitudes using also a combination of \g-ray lines. Moreover, our results are also at the level of previous works on MC simulations of Compton cameras applied to PGI~\cite{Ortega:15, Draeger:18, Munoz2021}, and better than the 5~mm obtained in Ref.\cite{McCleskey:15} or the 7~mm reported in Ref.~\cite{Taya:16}. 

The i-TED detector was developed for neutron-capture time-of-flight experiments, where a low sensitivity to neutrons is a crucial aspect. Neutrons are also among the main contributors to the secondary patient dose and to the background in PGI systems. Indeed, previous works have discussed the possible rejection of the neutron background in PGI systems through the application of time-of-flight selections~\cite{Biegun:12,Cambraia:18,Fontana:20}. However, aiming at a maximum detection efficiency, the detector-to-phantom distance has to be minimized and a satisfactory TOF separation of the prompt gammas and neutrons is not feasible (see Fig.~\ref{fig:nsensitivity}). In this scenario, the use of scintillators with reduced neutron sensitivity, such as LaCl$_{3}$ or BGO, is the best solution to minimize the neutron-induced background. Still, LaCl$_3$ shows the advantage of the high energy resolution for Compton imaging.

The maximum attainable efficiency of the full i-TED setup (see Table~\ref{tab:efficiency}) in the energy range of interest for PGI (1-7~MeV) is at the level of the most efficient existing or designed Compton cameras for PGI~\cite{Yao:19}, which reports an efficiency of 4.1$\times$10$^{-4}$ per proton for deposited energies above 1~MeV. The absolute efficiency per emitted prompt \g-ray for the i-TED array, calculated as the average of the four main PG transitions, is in the range 0.9-1.6$\times$10$^{-3}$ for the focal distances of Table~\ref{tab:efficiency}. This value outperforms most of the CC for PGI developed to date, with a range of efficiencies between 10$^{-4}$ and 10$^{-8}$ per emitted PG~\cite{Krimmer:18}. The high efficiency for the proposed setup over the whole energy range of interest (1-7~MeV) allows to determine the position of the PG distribution fall-off with an accuracy better than 3~mm for proton intensities as low as 10$^{8}$, in the range of clinical interest~\cite{Hueso:16,Smeets:12,Draeger:18}.  

Following the promising prospects for range verification in proton therapy with i-TED presented in this work, we aim at testing this detection system in proton beam facilities, such as the 18~MeV cyclotron at CNA (Sevilla) and clinical beams. These tests will provide an experimental validation for the methods and results presented in this work and they will allow us to explore possible additional aspects that can be optimized for the present ion-range monitor application.

\section*{Methods}\label{sec:Methods}

\subsection*{The i-TED Compton imager}
This work has presented a MC study based on the i-TED array consisting of four high-efficiency Compton cameras~\cite{Domingo:16,Babiano:21}. This novel detection system is under development at Instituto de F\'isica Corpuscular (IFIC) within the HYMNS-ERC project~\cite{hymns}. The first demonstrator has been already assembled, fully characterized~\cite{Babiano:20} and applied to high-resolution neutron TOF experiments~\cite{Babiano:21}. 

\begin{figure}[!htbp]
\centering
\includegraphics[width=0.4\linewidth]{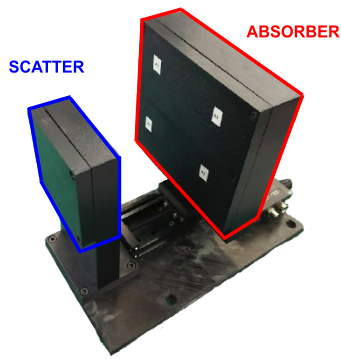}
\includegraphics[width=0.4\linewidth]{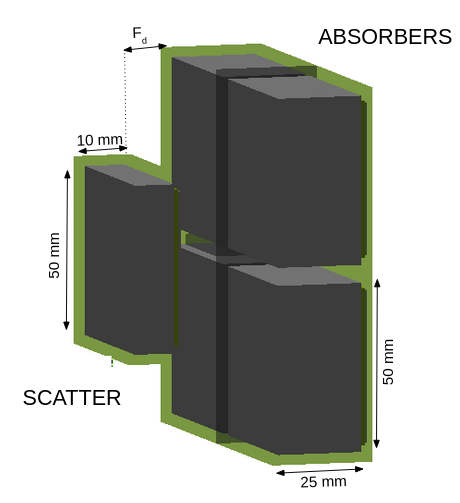}
\caption{Left: i-TED detector consisting of one scatter and four absorber detectors in movable and parallel detection planes. Right: Schematic view of the same i-TED detector as implemented in \textsc{Geant4} indicating the dimensions of the LaCl$_{3}$(Ce) crystals of the scatter and absorber planes.}
\label{fig:iTED5}
\end{figure}

 Each of the i-TED Compton modules uses 5 position-sensitive detectors (PSDs) distributed in two parallel detection planes, Scatter (S) and Absorber (A), as shown in Fig.~\ref{fig:iTED5}. Each PSD contains a LaCl$_{3}$(Ce) monolithic crystal with a square-cuboid shape and a base surface of 50$\times$50~mm$^2$. The LaCl$_{3}$ is hygroscopic and thus it is encapsulated in an aluminum housing. Each crystal base is coupled to a 2~mm thick quartz window, which is optically joined to a silicon photomultiplier (SiPM) from SensL (ArrayJ-60035-64P-PCB). The photosensor features 8$\times$8 pixels over a surface of 50$\times$50~mm$^2$. A 15~mm thick crystal is used for the PSD in the S-plane. Four 25~mm thick crystals are utilized for the PSDs placed in the A-plane (see Fig.~\ref{fig:iTED5}). In total, 320 SiPM channels are biased and readout by means of front-end and processing PETsys TOFPET2 ASIC electronics~\cite{PETsys16}. In order to minimize gain shifts due to changes in the temperature of the experimental hall, every ASIC is thermally coupled to a refrigeration system composed by a Peltier cell, a heat-sink and a small-size fan (see Ref.~\cite{Babiano:20} for further details). 

The i-TED Compton modules embed the so-called dynamic electronic collimation technique~\cite{Babiano:20}. This is accomplished by means of a linear positioning stage, that allows one to remotely vary the distance between the A- and S-planes, thereby optimizing performance for each specific application. Finally, the successful implementation of ASIC-based TOF-PET readout electronics for Compton imaging has led to a rather compact and cost-effective system, when compared to other Compton imagers~\cite{Kataoka:2013,Nagao:2018}.

\subsection*{MC simulations of the proton beam and i-TED}

The applicability of i-TED for range verification has been studied by means of MC simulations using the \textsc{Geant4} toolkit. The detailed geometry model of each of the i-TED detectors as implemented in \textsc{Geant4} is shown in Figure~\ref{fig:iTED5}. More details can be found in our previous work~\cite{Babiano:21}. 

\begin{figure}[!htbp]
\centering
\includegraphics[width=0.45\linewidth]{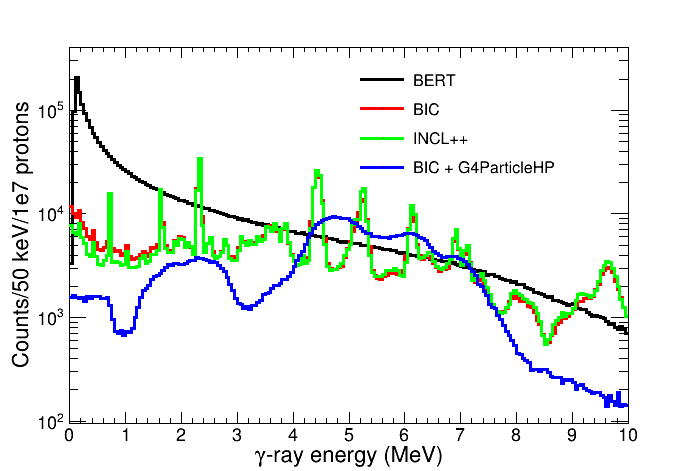}
\includegraphics[width=0.45\linewidth]{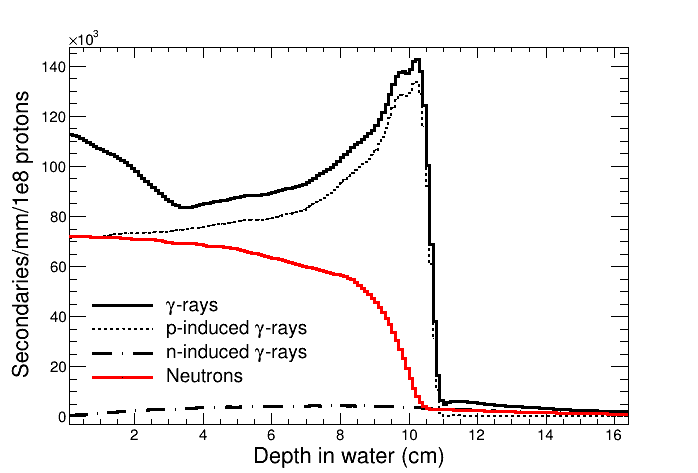}
\caption{Left: Energy spectrum of proton-induced prompt \g-rays obtained in the \textsc{Geant4} simulations with different Physics Lists. Right: Distribution of \g-rays (solid black) and neutrons (red) produced by 120~MeV protons as a function of the depth in water simulated with \textsc{QGSP\_INCLXX\_HP}.}
\label{fig:PG_MC}
\end{figure}

The modelling in \textsc{Geant4} of the physics processes occurring during the irradiation of a phantom with protons can be carried out with different models, so-called Physics Lists (PL)~\cite{Geant4PL}. In this work we have tested several officially released PLs, which combine the Quark-Gluon-String model (QGSP) for the inelastic scattering of protons above $\sim$10~GeV, not relevant for the present study, with three different cascade models covering the energy below 10~GeV: the Li\'ege Intranuclear Cascade model INCL++, the Bertini (BERT) or the Binary Cascade model (BIC). The resulting prompt gamma yields for the different PLs, presented in Fig.~\ref{fig:PG_MC}, indicate that INCL++ and BIC agree within 2\% in the absolute PG yield while BERT leads to a 2.3 times higher production, in agreement with the clear overestimation reported in Ref.~\cite{Pinto:16}. Moreover, the BERT model generates a continuum of \g-ray energies instead of the expected discrete spectra (see Fig.~\ref{fig:PG_MC}), as it has been reported in previous works~\cite{Verburg:12}.  Last, the \textsc{QGSP\_BIC\_AllHP} Physics List was tested. This PL includes the G4ParticleHP package, still under development, which uses nuclear data libraries instead of the default models for the transport of light charged particles~\cite{G4ParticleHP}. In particular, the inelastic interaction of protons with $^{12}$C and $^{16}$0 up to 150~MeV are simulated using the ENDF/B-VII.1~\cite{ENDF} cross section. This PL leads to a 32\% smaller PG yield, a result which is in line with the reduction suggested by the benchmark of Pinto et al.~\cite{Pinto:16}. However, the latter was discarded since it failed to produce individual \g-ray lines in the proton-induced inelastic reaction (see Fig.~\ref{fig:PG_MC}).

As for the neutron yields in the irradiated phantoms, all the studied PLs agree within 24\%. For an accurate transport of neutrons below 20~MeV, neutron-induced reactions within \textsc{Geant4} were simulated by means of the G4NeutronHP package~\cite{Mendoza:14}, using the G4NDL-4.6 data library (based on the JEFF-3.3~\cite{Plompen:20} evaluated data file). This high-accuracy neutron transport package, not included in some of the previous works~\cite{McCleskey:15,Yao:19}, has a significant impact in the slowing-down and partial capture of neutrons within the phantoms, and in the response of the detectors to neutrons. Indeed, the number of slow neutron events in the detectors would have been underestimated in up to a factor 5 if this package had not been included.

The final choice of PL was \textsc{QGSP\_INCLXX\_HP} since it includes an accurate modelling of neutron interactions and leads to the smallest PG yield among the PLs, which correctly generate discrete \g-ray transitions. In addition, the INCL++ model has shown the best reproduction of the neutron and \g-ray yields in the simulation of proton-induced reactions in other applications~\cite{LoMeo:15,Lerendegui:16_2}. The right panel of Figure~\ref{fig:PG_MC} shows the depth profile for the secondary emission of \g-rays and neutrons in water resulting from our MC simulations of the 120~MeV proton beam. In the case of the \g-rays, the largest production is due to proton-induced reactions (dotted line), which show a clear maximum at the end of the proton range. The much smaller neutron-induced production of \g-rays is shown with the a dotted-dashed line in the same figure. For the PG images obtained in this work, several energy cuts have been applied to select either the whole spectrum of Fig~\ref{fig:PGSpectra} between 1-7~MeV or the main single transitions.

\begin{figure}[!htbp]
\centering
\includegraphics[width=0.43\linewidth]{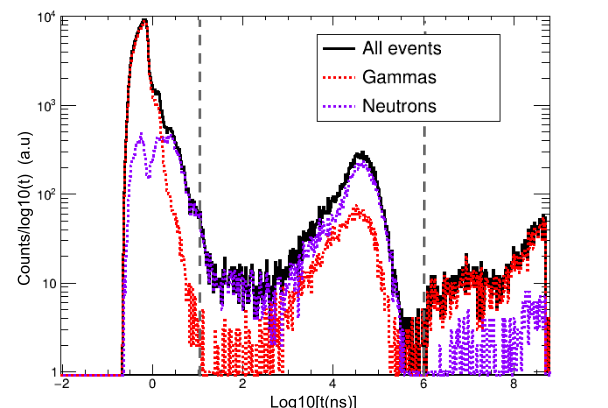}
\includegraphics[width=0.47\linewidth]{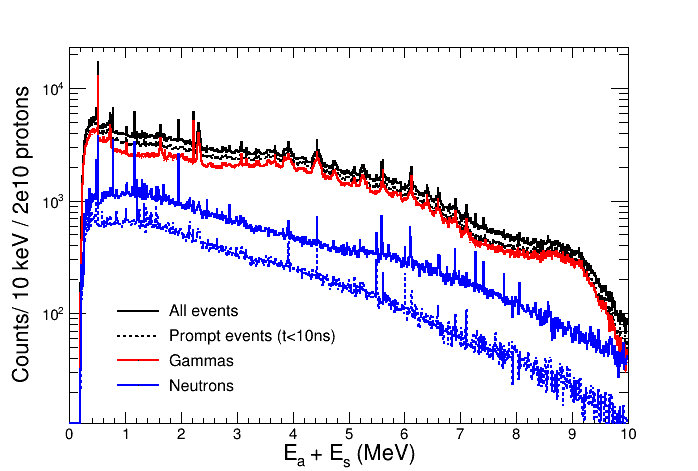}
\caption{Left: time distribution (log scale) of the \g-ray and neutron events registered in i-TED, showing three distinct components which are separated by vertical dashed lines (see text for details). Right: add-back energy spectrum for all and prompt (t$<$10~ns) events .}
\label{fig:time_energy_G_N}
\end{figure}

A particular emphasis has been placed on the physical origin and time distributions of the \g-rays and neutrons escaping the phantom and reaching the i-TED detectors. This is of particular relevance for our study of the neutron sensitivity and the applicability of time-of-flight cuts. The left panel of Fig.~\ref{fig:time_energy_G_N} shows the time of arrival of \g-rays and neutrons to i-TED after being produced in a water phantom by 120~MeV protons. Three time structures can be identified in this figure, whose limits are indicated with dotted lines. The prompt particles, \g-rays in their majority, reach i-TED within the first 10 ns. A second component, which extends up to 1~ms after the proton bunch, contains both neutrons moderated in the phantom and \g-rays originated in neutron capture reactions, dominated by the 2.2~MeV produced in the $^{1}$H(n,\g) reaction. Beyond 1~ms, the third emission \g-ray component is related to the decay of the unstable nuclei produced in the phantom, simulated with the G4RadioactiveDecay model included by default. To avoid the full simulation of the decays, the transport of particles was limited to 1~ms. The add-back spectra of i-TED operated in coincidences is shown in the right panel of Fig.~\ref{fig:time_energy_G_N}. This figure shows the strong reduction of the neutron sensitivity when a time cut of 10~ns is applied. The integral of the gamma- and neutron-induced spectra of Fig.~\ref{fig:time_energy_G_N} correspond to the results of i-TED (LaCl$_{3}$) presented in Fig.~\ref{fig:nsensitivity}.

\subsection*{Compton imaging algorithms}
Effective prompt gamma ray monitoring requires images with spatial resolutions down to a few mm and reconstruction times of the order of a few seconds, at most. Compton imaging uses as a fundamental ingredient the Compton scattering law and thus, most Compton cameras employ two layers, or more, of position sensitive detection planes. The direction of the scattered \g-ray is determined from the interaction position in both detection planes. In case that the scattered \g-ray deposits all its energy in the second detection plane, the incident \g-ray energy can be determined by adding-up the deposited energy in the first and second detection planes. The Compton imaging algorithms used in this work are based on this assumption, whose validity is extended by means of a supplementary machine learning classifier, that will be discussed in the following section. 

There exist many different imaging reconstruction algorithms, each of them with advantages and drawbacks. For instance, algorithms based on Maximum Likelihood Expectation Maximization~\cite{Wilderman:98-2,Kim:10} have shown excellent position reconstruction accuracy, but they require previous computation of the system response matrix. Since the aim of this work is to study the applicability of an imager like i-TED for clinical purposes, we have chosen algorithms which do not require any previous system response calculation. Those algorithms are listed as follow:

\begin{itemize}
    \item {\bf{Fast back-projection (BP)}}~\cite{Wilderman:98-1}: This is the simplest approach, with rather limited spatial resolution, but also with the fastest reconstruction performance. Developed by Wilderman in 1998, the image of the individual \g-rays is obtained from the intersection of the back-projected Compton cones along the image plane, where the \g-ray source is located. The general quadratic curve from this intersection leads to a set of possible positions for the \g-ray source origin in the image plane. The final Compton image is made by the superposition of all the individual \g-ray images. While the algorithm is very fast, the spatial resolution is quite poor, when compared to other algorithms (see for example Fig~\ref{fig:2d all}).   
    
    \item {\bf{Stochastic origin ensemble (SOE)}}~\cite{Andreyev:16}: The SOE algorithm for Compton imaging was developed originally by Andreyev in 2016. It is a Monte Carlo Markov chain Method based on the Metropolis-Hasting algorithm and does not require any forward or backward projections. The drawback of this algorithm is related to its iterative nature. The initial image, obtained from all the statistics available, is generated by random sampling of the possible positions of the \g-ray source within the intersection of Compton conical surfaces and the image plane. Once this first image is created, the iterative process begins by randomly choosing a \g-ray event. A new position for this event is sorted in the image space constrained to its Compton conical surface. Then, this new position is either accepted or rejected with an acceptance $A$ based on the ratio between the local density of \g-ray events in the new ($\lambda^{\prime}_{i}$) and old ($\lambda_{i}$) positions:
    \begin{equation}
        A=min\left(1,\frac{\lambda^{\prime}_{i}+1}{\lambda_{i}}\right)
    \end{equation}
    This is repeated for a number of times equivalent to the available statistics. This process corresponds to an iteration and it is repeated until a stationary situation is reached. The number of iterations required depends on the specific problem, such as the available statistics and the binning of the Compton image. 
    
    \item {\bf{Analytical algorithm (AA)}}~\cite{Tomitani:02}: An analytical inversion of the Compton imaging problem based on spherical harmonics was developed by Tomotani and Hisarawa in 2002. The approximate solution given a position in the image space, $\vec{s}$, is described as:
    \begin{equation}
        f(\vec{s})\approx\int^{cos\omega_{max}}_{cos\omega_{min}}{dcos\omega}\int_{S}{d\vec{t} k^{-1}(\vec{t},\vec{p};cos\omega)g(\vec{t};cos\omega)}
    \end{equation}
    where $\vec{t}$ is a unit vector into the projection space, $\omega_{min}$ and $\omega_{max}$ are the minimum and maximum Compton scattering angles included in the calculation, $g(\vec{t};cos\omega)$ is the projection data in the image space and $k^{-1}(\vec{t},\vec{p};cos(\omega))$ is the inversion kernel described as
    \begin{equation*}
        k^{-1}(\vec{t},\vec{p};\omega)=\sum^{N_{max}}_{n=0}{\frac{2n+1}{4\pi H_{n}}P_{n}(cos\omega)P_{n}(\vec{s}\cdot\vec{t})}
    \end{equation*}
    with $H_n$ given by
    \begin{equation}
        H_{n}=\int^{cos\omega_{max}}_{cos\omega_{min}}{\sigma(cos\omega)P^{2}_{n}(cos\omega)dcos\omega}
    \end{equation}
    $N_{max}$ is the maximum number of terms involved in the calculation, $P_{n}$ is the Legendre polynomial of order $n$ and $\sigma(cos\omega)$ is the Klein-Nishina Compton differential cross-section~\cite{Klein:29}. The drawback of this algorithm is the large computational cost required to reconstruct a Compton image. Because of this reason, $H_{n}$ was pre-computed for a wide range of \g-ray energies and Compton scattering angles corresponding to the minimum and maximum values can be experimentally registered by the detection system. Thus, $H_{n}$ values were saved in a table format for its posterior use, saving a crucial time for quasi-real time imaging.
\end{itemize}

The reconstruction time depends remarkably on the complexity of the involved algorithms and the optimization of their parameters such as the number of iterations for the SOE algorithm or the number of terms involved for AA reconstruction. Additionally, those parameters will impact the quality of the Compton image, introducing a mismatch between the reconstructed and the experimental spatial image resolution of the Compton imager; a limited number of iterations or terms in these imaging algorithms would lead to a poor spatial resolution, while an excessive number would introduce artifacts on the final reconstructed image.

In order to evaluate the best suitable algorithm for PG monitoring, a comparison benchmark using 20000 \g-ray events leading to interactions in both layers were used. The image plane of 200$\times$200~mm$^2$ was pixelated into 200$\times$200 pixels for all three algorithms. The results for the computing time obtained in this benchmark study are displayed in Table~\ref{tab:Comp_time}. 

\begin{table}[ht]
\centering
\begin{tabular}{|l|l|l|l|}
\hline
Algorithm & Implementation & Parameters &  Time (s) \\
\hline
BP & (CPU Single-thread) & None & $<$5 \\
\hline
SOE & (CPU Single-thread) & $n_{iter}$ = 1000 & 14  \\
\hline
AA & (CPU Single-thread) & $N_{max}$ = 70 &  1821 \\
\hline
AA & (CPU Multi-thread 8) & $N_{max}$ = 70 & 260   \\
\hline
AA & (GPU-CUDA) & $N_{max}$ = 70 & 15  \\
\hline
\end{tabular}
\caption{\label{tab:time} Reconstruction time for a Compton image of 20000 events. First and second columns show the type of algorithm and its implementation in CPU or GPU. The relevant parameters for the SOE and AA algorithms are indicated in the third column (see text for details). }
\label{tab:Comp_time}
\end{table}

The critical parameters for each algorithm used in the benchmark are indicated in the third column of the same table. Their values correspond to the best compromise between resolution and computation speed found by trial and error. As the results indicate, the fastest algorithm is BP followed by SOE and AA, being a factor 100 slower than BP. However, given the superior spatial resolution of AA, both multi-threading and CUDA~\cite{CUDA} implementations were used to make this algorithm time competitive with the others. Compared to the single-thread version, the multi-thread implementation, with a total of eight threads, has a speed factor of about 7.6. Despite the improvement, the time required to get the image is still not competitive for PG imaging with a factor between 17 and 30 times slower than the other algorithms. On the contrary, the CUDA version, with a speed factor of about 121 with respect to the singled-threaded version, reaches reconstruction times of 15~s, comparable to the SOE algorithm (14~s). It is worth to mention that the CUDA implementation was executed in modern NVIDIA GPUs: a NVIDIA GeForce RTX 2060 and NVIDIA GTX 1080 Ti with compatible execution times.

\subsection*{Machine-learning aided full-energy event selection}
The i-TED detector, as any two-plane Compton camera, bases its working principle on a \g-ray totally depositing its energy  in the absorber (A) plane after a Compton interaction in the scatter (S) layer. However, only a fraction of the coincidences, ranging from 47\% for 1~MeV \g-rays to just 5\% at energies around 7~MeV, satisfy this condition. The remaining fraction leads to a wrong reconstruction of the Compton angle and, as a consequence, an increase in the image background and a degradation of the spatial resolution. Aiming at improving the fraction of true Compton events, we have implemented a Machine-Learning algorithm for the identification of full-energy events.

To train and test the ML algorithms in the discrimination of good Compton events where the \g-ray deposits all its energy between the S and A planes (full-energy), and those with partial energy escape (no-full-energy), we performed dedicated simulations of the response of the i-TED detectors to 5$\times$10$^{10}$ \g-rays of energies homogeneously distributed between 200 keV and 7 MeV and spatially originated in a random position within a 20$\times$20$\times$20~cm air cube separated 50~mm from the detector face, replicating the position of the phantom. For these ancillary simulations, we used the same physics models and distance between the S- and A-planes of i-TED as before. Each MC event (S\&A coincidence) contains the same eight features determined with the detector in a real measurement: 3D coordinates of the \g-ray interactions in the two PSDs (6), energy deposited in the S- and A-planes (E$_{s}$, E$_{a}$) (2). The energy and position resolutions of the detector were included as described before. Additionally, the Compton angle, calculated from the deposited energies, and its probability according to the Klein-Nishina formula~\cite{Klein:29} for a \g-ray energy E$_{\gamma}$=E$_{a}$+E$_{s}$, were also included in the training to improve performance. The MC output was split into 14 energy intervals of add-back deposited energy between 200~keV and 7~MeV and the same number ($\sim$10$^{6}$) of either kind of events were selected from the MC output for each energy interval. For each energy range we trained the same algorithm independently, aiming for the best accuracy and stability along the entire deposited energy range detected in add-back. 

In this work, the performance of several state-of-the-art ML algorithms included in \textit{Scikit-learn} Python module~\cite{scikit-learn} was evaluated. The \textit{Scikit-learn} algorithms evaluated were k-nearest neighbors, logistic regression, support vector classifier, Gaussian naive Bayes, random forest, AdaBoost and quadratic discriminant analysis. However, the best classification results in terms of accuracy were obtained for other two ML algorithms: Boosted decision trees implemented in XGBoost~\cite{Chen:16} and artificial neural networks implemented in Tensorflow~\cite{tensorflow:15,chollet:15}.

\begin{figure}[!htbp]
\centering
\includegraphics[width=0.45\linewidth]{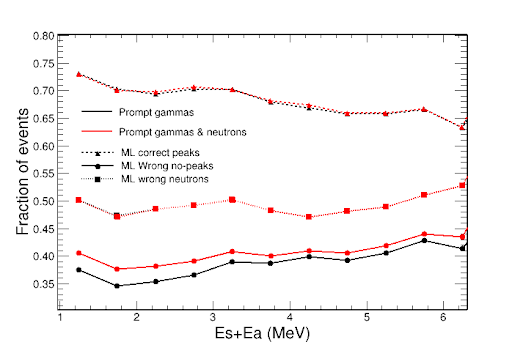}
\includegraphics[width=0.45\linewidth]{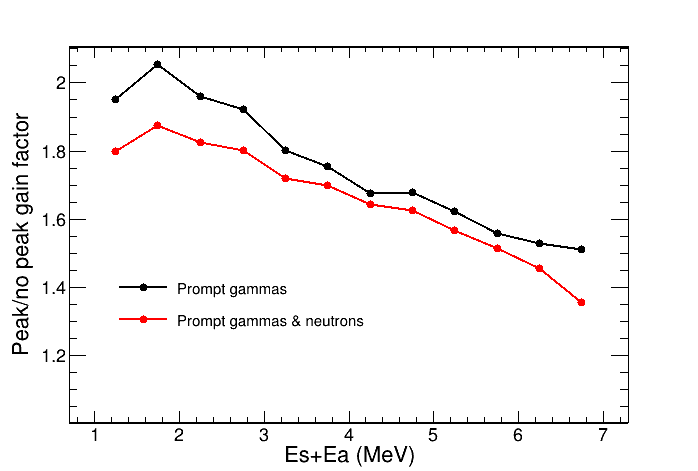}
\caption{Left: Fraction of correctly identified full-energy events (dashed) and wrongly classified no-full-energy (solid) as a function of the add-back deposited energy. The curve with squares correspond to the fraction of neutron events which are identified by the ML-classifier as good (full-energy) events.  Right:   gain factor of full- over no-full-energy events as a function of the add-back deposited energy. In both panels, the black and red lines represent, respectively, the results with only \g-ray events and after including the neutron events.}
\label{fig:MLClassifier}
\end{figure}
       
The XGBoost classifier was optimized using the \textit{Scikit-learn} GridSearchCV method~\cite{scikit-learn}. The best results were obtained using 140 trees with a depth of 4 and a gamma parameter of 0.1. The rest of parameters were set to default because of the minimal impact on the accuracy results. The classifier output is given by a probability number between 0 (no-full-energy event) and 1 (full-energy event). The discrimination threshold between classes was set to 0.5. The Tensorflow classifier model consists of a stack of four fully connected layers. In the first three layers, 256 units per layer using \textit{rectified linear} activation functions. In the last layer, a single neuron with a \textit{sigmoid} activation function was used with the purpose to provide a probability number between 0 and 1. The architecture and activation function was chosen based on the best performance in terms of accuracy and stability along the entire range of deposited energy in add-back. The loss function used for the minimization was \textit{binary cross-entropy}, commonly used in this type of classification problem because of its best performance. As in the case of XGBoost, the identification threshold was set to 0.5. 

XGboost and Tensorflow models have almost the same performance, being the latter slightly more stable for all the different deposited energy ranges where the training and test was performed. For this reason, only the model based on Tensorflow was used for the final results of this work. The accuracy of this algorithm was quantified on i-TED events from the simulations of the proton beam prior to its application to the image reconstruction. As shown in Fig.~\ref{fig:MLClassifier}, the ML algorithm is able to correctly recognize 65 to 73\% of the full-energy events among the prompt gamma events registered in i-TED. On the other hand, 35 to 40\% of the non-peak \g-ray events are wrongly predicted as full-energy. As a combination of both results, this algorithm enhances the peak-to-background ratio by a factor of 1.5-2.1 (see right panel of Fig.~\ref{fig:MLClassifier}). This classifier is also able to reject about 50\% of the neutron events. The final peak-to-total gain factor, after the neutron events are included, ranges from 1.4 to 1.9 (see Fig.~\ref{fig:MLClassifier}).


\section*{Acknowledgements}
This work has been carried out in the framework of a project funded by the European Research Council (ERC) under the European Union's Horizon 2020 research and innovation programme (ERC Consolidator Grant project HYMNS, with grant agreement No.~681740). The authors acknowledge support from the Spanish Ministerio de Ciencia e Innovaci\'on under grants PID2019-104714GB-C21, FPA2017-83946-C2-1-P, FIS2015-71688-ERC, CSIC for funding PIE-201750I26.

\section*{Author contributions statement}
\textbf{J.L.M.} Investigation, Methodology, Formal analysis, Data curation, Visualization, Writing - original draft. 
\textbf{J.B.C:} Investigation, Methodology, Software, Data curation, Writing - original draft.
\textbf{V.B.S} Investigation, Methodology.
\textbf{I.L:} Software, Visualization.
\textbf{C.D.P:} Conceptualization, Methodology, Supervision, Writing -review \& editing, Funding acquisition.

\section*{Additional information}
\textbf{Competing interests:} The authors declare that they have no known competing financial interests or personal relationships that could have appeared to influence the work reported in this paper.

\end{document}